\documentclass{ws-ijmpd}
\usepackage{amsmath}
\usepackage{amsfonts}
\usepackage{amssymb}
\usepackage{mathrsfs}
\usepackage{color}

\def\bc{\begin{center}}
\def\ec{\end{center}}
\def\nno{\nonumber}
\def\be{\begin{eqnarray}}
\def\ee{\end{eqnarray}}

\newcommand{\omits}[1]{}

\definecolor{dyellow}{rgb}{1.,0.8,.0}
\definecolor{myblue}{rgb}{.1,.1,.7}
\definecolor{dcyan}{rgb}{.0,.6,.6}
\definecolor{dmagenta}{rgb}{0.6,0.0,0.6}
\definecolor{brown}{rgb}{0.6,0.2,0.}
\definecolor{darkblue}{rgb}{.0,.0,0.5}
\definecolor{darkred}{rgb}{0.75,0.0,0.0}
\definecolor{orange}{rgb}{1.,.6,.0}
\definecolor{dorange}{rgb}{0.8,.4,.0}
\definecolor{darkgreen}{rgb}{0.0,0.6,0.0}
\definecolor{purple}{rgb}{.4,.0,.4}

\def\Ga{\Gamma}

\def\dl{\delta}

\def\la{\lambda}

\def\si{\sigma}
\def\om{\omega}

\def\del{\nabla}
\def\d#1#2{\frac{\displaystyle #1}{\displaystyle #2}}
\def\r{\partial}

\newcommand\btd{\raise 2pt
\hbox{$\hat\bigtriangledown$}\hskip 1.5pt}
\newcommand\bt{\raise 2pt
\hbox{$\bigtriangledown$}\hskip 1.5pt}

\begin{document}

\title{A NEW KIND OF UNIFORMLY ACCELERATED REFERENCE FRAMES}

\author{Chao-Guang HUANG}

\address{Institute of High Energy Physics, Chinese Academy of
Sciences\\
P.O. Box 918-4, Beijing 100049, P. R. China\\
huangcg@ihep.ac.cn}
\author{Han-Ying GUO}
\address{Institute of Theoretical Physics,
 Chinese Academy of Sciences \\
 P.O.Box 2735, Beijing 100080, P. R. China \\
hyguo@itp.ac.cn}

\maketitle

\begin{abstract}
{A new kind of uniformly accelerated reference frames with a
line-element different from the M\o ller and Rindler ones is presented, in which
every observer at $x, y, z=$consts. has the same constant
acceleration. The laws of mechanics are checked in the new kind of
frames.  Its thermal property is studied.  The comparison with the
M{\o}ller and Rindler uniform accelerated reference frames is also
made.}
\end{abstract}

\bigskip



\date{March 9, 2005}

%



\section{Introduction}

It is very well-known that an inertial reference frame in a flat
spacetime equips with a Minkowski coordinates so that
\be \label{Mink}
ds^2=c^2dt^2-dx^2-dy^2-dz^2.
\ee

In the literature, a uniformly accelerated reference frame is
defined as a set of observers who remain at rest with respect to
a given observer {\it Alice} who is accelerating at a constant
rate with respect to the instantaneously comoving inertial frames\cite{DP}.  Two observers {\it Alice} and {\it Bob} are said to be
{\it at rest} with respect to each other if the time elapses
on the clock of {\it Alice} during the passage of a light signal
from {\it Alice} to {\it Bob} and back again always has the same value. By definition, two observers who at each instant are
moving with the same velocity $v(t)$ with respect to an inertial
frame will remain at rest only when $v(t)$ is a constant.
Therefore, a set of observers who has a single acceleration
is not said to constitute a uniformly accelerated reference frame.
Instead, a uniformly accelerated reference frame is usually
described by
Rindler coordinate system\cite{Rindler77}
\be \label{Rindler}
ds^2=  {\tilde\xi}^2  d{\tilde\eta}^2 - d{\tilde\xi}^2 - dy^2 -dz^2,
\ee
or M{\o}ller coordinate system\cite{Moller}\cdash\cite{Gron}
\be \label{Moller}
ds ^2=(1+g \xi)^2d \eta^2 -d \xi^2- dy^2 -dz^2,
\ee
in which static observers at different places in the spacetime, whose spatial coordinates are constants, have different accelerations.

It should be noted, however, that such a kind of uniformly
accelerated reference frame is
not suitable to describe the frame consisting of a group of
electrons in a uniform electric field or the motion of phonon
in an accelerated crystalloid.

On the other hand, when a clock experiences a uniform acceleration, its
time elapses at different rate time to time.  Thus, to define ``{\it at
rest}" with the help of such a kind of clock is not without questions.
The concept `{\it at rest}' of such a kind is quite different from
that in the laboratory (inertial) frame.  Obviously, an alternative
definition of `{\it at rest}' will lead to an
alternative definition of a uniformly accelerated reference frame
and thus an alternative, uniformly accelerated coordinate system.

In the present paper, we shall
study a new kind of reference frame in which each observer moves
with the same acceleration, which is thought to be unable to serve as an
acceleration reference frame in Ref. \refcite{DP}.  Since the reference
frame have only constant parameter $a$ distinguishing from an
inertial reference frame, we also refer to it as a uniformly
accelerated reference frame.

 The paper is organized as
follows. In Sec. \ref{Sec:Rindler}, the Rindler spacetime is
reviewed briefly. In Sec. \ref{AccRef}, uniformly accelerated
reference frames in flat spacetime are introduced.  Sec. \ref{temperature}
is a short discussion on the thermal property of the new
uniformly accelerated reference frame.  In the
last section, the concluding remarks are given.


\section{Rindler Spacetime}\label{Sec:Rindler}

In Rindler spacetime (\ref{Rindler}), the 4-velocity of a
particle standing at constant-${{\tilde\xi}, y, z}$ is
$U^\mu=\{{\tilde\xi} ^{-1}, 0, 0, 0\}$ and its 4-acceleration
is
\be
a^\mu &:=& U^\nu \del_\nu U^\mu =\{0, {\tilde\xi}^{-1}, 0, 0\}.
\ee
The magnitude of the acceleration is $a=(-g_{\mu \nu}a^\mu a^\nu)^{1/2}={\tilde\xi}^{-1}$, which depends on the position of particle in the spacetime.
Let ${\tilde\eta} =g \eta$ and ${\tilde\xi} = (1+g \xi)/g$, where
$g$ is a constant.  Then the line-element (\ref{Rindler}) is
written in the M\o ller coordinates (\ref{Moller}).
The magnitude of the acceleration of a particle standing at
constant-${\xi, y, z}$ is
\be \label{acc-nu}
a=\d g {1+g \xi}.
\ee
In particular, $a=g$ at $ \xi =0$.
The horizon for (\ref{Moller}) is at $ \xi_H =-1/g$.
By definition, the ``surface gravity" on the horizon is\cite{wald}
\be
\kappa := \lim_{\xi \to  \xi_H } (Va) = g,
\ee
where $V=\sqrt{g_{00}}=1+g \xi$ is the redshift factor, and
the Hawking temperature is
\be \label{T_H-nu}
T_H:=\d \kappa {2\pi} =\d g {2\pi}.
\ee
All above equations are rewritten in natural units.  In SI units
which is used in the following discussion,
the line-elements (\ref{Rindler}) and (\ref{Moller}), acceleration (\ref{acc-nu}), and Hawking temperature (\ref{T_H-nu}) may be written as
\be \label{Rindler1}
ds^2=  \d {c^5}{G_N\hbar} \tilde \xi^2  d\tilde \eta^2 - d\tilde \xi^2 - dy^2 -dz^2
\ee
and
\be \label{Rindler3}
ds ^2=\left ( 1+ \d {g\xi} {c^2} \right )^2 c^2d \eta^2 -d \xi^2 - dy^2 - dz^2,
\ee
\be \label{acc}
a=\d g {1+g \xi/c^2},
\ee
and
\be \label{T_H}
T_H =\d {g\hbar} {2\pi ck_B},
\ee
respectively.  It is obvious that the line-element (\ref{Rindler3})
is more natural to describe an accelerated observer than
Eq.(\ref{Rindler1}) because Eq.(\ref{Rindler3}) only contains the
speed of light $c$ while Eq.(\ref{Rindler1}) also contains the
Newtonian gravitational constant $G_N$ and Planck constant $\hbar$
which are not related to accelerated motion directly.

For metrics (\ref{Rindler}) and (\ref{Moller}), the congruence of trajectories of accelerated observers has vanishing expansion,
spatial components of shear and twist\cite{SachsWu}.  The non-zero
components of shear and twist are only $\si^{01}=\si^{10} =
-\frac 1 2 \r_1 U^0$ and $\om^{01}=-\om^{10} =\frac 3 2 \r_1 U^0$.
Therefore, the distance between two static observers, {\it Alice} and
{\it Bob}, measured in the comoving inertial frame always has the same
value.


\section{New Uniformly Accelerated Reference Frame and Its Line-element in Flat Spacetime }\label{AccRef}

Now, we consider a uniformly accelerated reference frame consisting
of a set of accelerated observers with proper acceleration $a$ in a
given direction, say, in $x-$direction, as shown in Fig. 1.  At $t= 0 $, the observers are at $(x=X,y,z)$,
where $t,x,y,z$ are Minkowski coordinates of a flat spacetime.   The
trajectory of each observer is a hyperbola of same eccentricity but
different asymptotes in the Minkowski spacetime.  Namely, different
observers have different horizons.
\begin{figure}[hbt] \label{FIG:arfig}
\centerline{\includegraphics[scale=0.8]{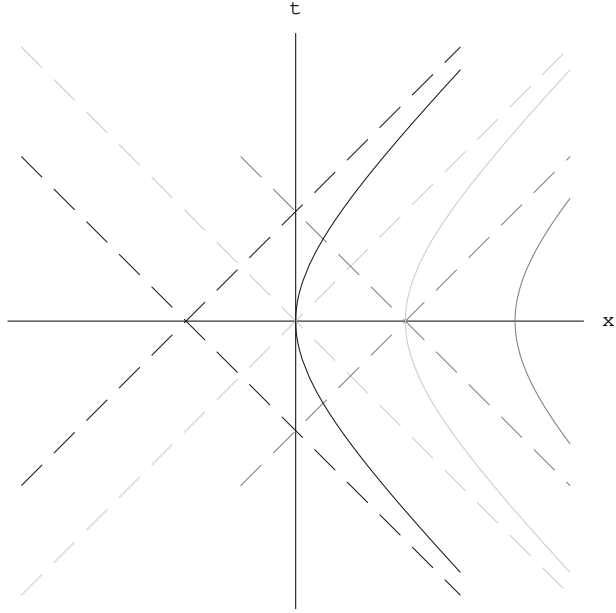}}
\vspace*{8pt}
\caption{Trajectories (real) of observers in an accelerated reference
frame.  Different observers 
in the same accelerated reference frame have different horizons (dash).}
\end{figure}

The trajectory of an accelerated observer with acceleration $a$
passing $(0,X,y,z)$ at $t=0$ is (see, e.g.\cite{Rindler77})
\be \label{traj}
x(t)=X+\d {c^2}{a}\left [ \left(1 + \d {a^2t^2}{c^2}  \right )^{1/2}-1\right ].
\ee
Since
\be
\d {dt} {ds} &=& \frac 1 c\frac {x(t)-X+\frac {c^2}{a}} { [ (x(t)-X+\frac{c^2}{a} )^2- c^2t^2  ]^{1/2}} \nno  \\
\d {dx} {ds} &=&\frac {ct} {[(x(t)-X+\frac{c^2}{a})^2- c^2t^2 ]^{1/2}}\\
\d {dy} {ds} &=& \d {dz} {ds} =0, \nno \ee
on the trajectory
\be
\d {dt} {ds} = \d 1 c \left[1+ \d
{a^2}{c^2}t^2 \right ]^{1/2}. \ee
Its integration under the initial
condition that $t=0$ when $s=0$ is
\be \label{t-s}
t=\d c a \sinh \d {a s}
{c^2}. \ee
Substituting it in Eq.(\ref{traj}), we obtain
\be \label{x-X}%
x(t)=X+\d {c^2} {a} \left (\cosh \d {a s} {c^2} -1\right ). \ee

In order to get a line-element for the uniformly accelerated
reference frame in the flat spacetime, we choose new coordinates
$(s=:cT, X, y, z)$ so that
\be
dt &=& \cosh \d {aT} c dT \label{2time}\\
dx &=& dX+ c \sinh \d {aT} c dT.
\ee
The line-element of the flat spacetime can be written as
\be \label{accref} ds^2 =c^2
d T^2 -2c\sinh \d {aT} c dTdX-dX^2-dy^2-dz^2
\ee
in terms of the new
coordinates.  This is the line-element for the uniformly
accelerated reference frame with the acceleration $a$.

To make sure that Eq.(\ref{accref}) describes an accelerated
reference frame, let us consider an arbitrary static observer
in the new coordinates such that
\be
U^\mu = c(1, 0, 0, 0).
\ee
The 4-acceleration
\be
a^\mu &=& U^\nu \del_\nu U^\mu =  a {\tanh \d {aT} c} \dl^\mu_0 + \d a {\cosh \d {aT} c}\dl^\mu_1
\ee
has the magnitude
\be
(-g_{\mu \nu}a^\mu a^\nu)^{1/2}=a,
\ee
which is independent of the position of a static observer in the spacetime.
This is unlike the Rindler spacetime in which the static observer
at different place has different acceleration.  Therefore, we say that
$a$ here is the acceleration of the accelerated reference frame and
that $(T,X,y,z)$ constitutes  an accelerated coordinate system.

It is remarkable that the accelerated coordinate
system differs from an inertial coordinate system only by a
non-diagonal term.  At the time origin, the concepts of space and time
in an accelerated reference frame are identical to those of inertial
observers.

It can be shown that such a kind of accelerating system is locally
synchronizable at $T=0$ hyperplane\cite{SachsWu}.

In an inertial reference frame, the second law of mechanics reads
\be
m_0 c \d {dU^\mu}{ds}=F^\mu .
\ee
Since the 4-velocity in an inertial reference frame, $U^\mu_{\rm iner}$, and that in the accelerated reference frame whose origin is at rest with
respect to the inertial reference frame at $t=0$, $U^\mu_{\rm acc}$, are related
by
\be
U^\mu_{\rm iner} =\d {\r x^\mu}{\r X^\nu} U^\nu_{\rm acc},
\ee
which reads
\be
U^0_{\rm iner} 
&=&U^0_{\rm acc}\cosh \d {aT} c  \nno \\
U^1_{\rm iner} 
&=&U^0_{\rm acc} \sinh \d {aT} c  + U^1_{\rm acc}\\
U^2_{\rm iner} &=&U^2_{\rm acc} \nno \\
U^3_{\rm iner} &=&U^3_{\rm acc}, \nno
\ee
\be
\d {d U^0_{\rm iner}}{d s} &= &\d {d U^0_{\rm acc}}{ds} \cosh \d {aT} c
+ \d a {c^3} (U^0_{\rm acc})^2 \sinh \d {aT} c \nno \\
\d {d U^1_{\rm iner}}{d s} &=& \d {d U^0_{\rm acc}}{ds} \sinh \d {aT} c + \d a {c^3} (U^0_{\rm acc})^2 \cosh \d {aT} c+\d {d U^1_{\rm acc}}{ds} \nno \\
\d {d U^2_{\rm iner}}{d s} &=& \d {d U^2_{\rm acc}}{ds} \\
\d {d U^3_{\rm iner}}{d s} &=& \d {d U^3_{\rm acc}}{ds}.\nno
\ee
Similarly,
\be
F^0_{\rm iner} &=&F^0_{\rm acc}\cosh \d {aT} c  \nno \\
F^1_{\rm iner} &=&F^0_{\rm acc} \sinh \d {aT} c  + F^1_{\rm acc}\\
F^2_{\rm iner} &=& F^2_{\rm acc}\nno \\
F^3_{\rm iner} &=& F^3_{\rm acc}.\nno
\ee
Then, the second law in the accelerated reference frame becomes
\be
m_0 c\d {dU^0_{\rm acc}}{ds}&=&F^0_{\rm acc}-m_0\d a {c^2} (U^0_{\rm acc})^2 \tanh \d {aT} c \nno \\
m_0 c\d {dU^1_{\rm acc}}{ds}
&=&F^1_{\rm acc} - m_0\d a {c^2} (U^0_{\rm acc})^2 {\rm sech} \d {aT} c \\
m_0 c\d {dU^2_{\rm acc}}{ds}&=&F^2_{\rm acc} \nno\\
m_0 c\d {dU^3_{\rm acc}}{ds}&=&F^3_{\rm acc}. \nno
\ee
It can also be obtained directly from
\be
F^\mu_{\rm acc} = m_0 c \d {D U^\mu_{\rm acc}} {ds},
\ee
since
\be
m_0 c \d {d U^\mu_{\rm acc}}{ds} &=& m_0 c\left ( \d {D U^\mu_{\rm acc}} {ds} - \d 1 c \Ga^\mu_{\nu \la}U^\nu_{\rm acc} U^\la_{\rm acc} \right), \nno \\
\Ga^\mu_{\nu \la} U^\nu_{\rm acc} U^\la_{\rm acc} &=&\d a {c^2} (U^0_{\rm acc})^2 (\tanh \d {aT}{c}  \dl^\mu_0 +{\rm sech} \d {aT}{c} \dl^\mu_1). \nno
\ee

In the Newtonian approximation,
$U^0_{\rm iner}\approx U^0_{\rm acc}\approx c$,
$c dU^0_{\rm iner}/(ds) \approx c dU^0_{\rm acc} /(ds) \approx 0$, $F^0_{\rm iner} \approx F^0_{\rm acc}\approx 0$, and $aT/c \to 0$,
the second law reduces to
\be
m_0 c\d {dU^i_{\rm acc}}{ds} =F^i_{\rm acc} - m_0 a \dl^i_1 .
\ee
It takes the standard form of the Newtonian mechanics in an accelerated reference frame.  In addition, in the Newtonian approximation, Eqs.(\ref{t-s}) and (\ref{traj})  reduce to
\be%
t=T, \qquad x=X+\d 1 2 a T^2.
\ee%
They are the standard relations between coordinates in an inertial and a
uniformly accelerated system.

In contrast, the second law of mechanics in the Rindler metric (\ref{Rindler1}),
\be
m_0 c \d {d U^\mu_{\rm R}} {ds} &=& F^\mu_{\rm R} - m_0 \Ga^\mu_{\nu \la}U^\nu_{\rm R} U^\la_{\rm R} \nno\\
&=&F^\mu _{\rm R}- m_0 U^0_{\rm R}[\tilde \xi^{-1} U^1_{\rm R} \dl^\mu_0 +\d {c^3} {G_N\hbar}\tilde \xi U^0_{\rm R} \dl^\mu_1]. \nno
\ee
reduces to
\be
m_0 c \d {d U^i_{\rm R}} {ds} =F^i _{\rm R}- \d {m_0 c^2}{\tilde \xi}   \dl^i_1
=F^i _{\rm R}- m_0 a(\tilde \xi)\dl^i_1. \nno
\ee
under the Newtonian approximation, where the approximation $g_{00}(U^0_{\rm R})^2 \approx c^2$ has been used.  It is not the standard form of
the second law Newtonian mechanics in an accelerated reference frame though the acceleration of constant-$\tilde \xi$ trajectory is $c^2/\tilde \xi$.  The second law of mechanics in the metric (\ref{Rindler3}) does not reduce to
the standard Newtonian form in an accelerated reference frame either.


\section{Wick Rotation of Uniformly Accelerated Coordinate System and
Temperature}\label{temperature}

Now, let us study the thermal property of the line-element (\ref{accref}) and consider its Euclidean section.  Under the Wick rotation %
\be %
T \to \tau = iT, \qquad s \to is_E^{} %
\ee 
the line-element (\ref{accref}) becomes %
\be %
ds_E^2 &=& d\tau^2 -2 c \sin \d {a\tau} c d\tau dX + dX^2 +dy^2+dz^2. %
\ee 
Clearly, the metric spacetime has a period in
$\tau$ direction. The period is $ 2 \pi c/a.$  Namely, the time variable in line-element (\ref{accref}) has the imaginary period $-i 2 \pi c/a.$  On the 
same reasoning as that in Ref.\refcite{GH}, the Green
function, as a function of coordinates, is expected to acquire the same
period automatically.  Therefore, one may expect, according to the Green's  function theory at finite temperature\cite{FTF},
that the observers in the accelerated
reference frame at uniform acceleration $a$ will observe the finite
Hawking temperature $a\hbar/(2\pi c k_B)$.  

To confirm that the 
observer will observe the finite Hawking temperature, more systematic 
studies on the Green's function are needed.  But it is beyond the scope of this paper.  Instead, it is recalled that the vacuum Green function for the detector with a uniform (proper) acceleration $a$ is the same as the thermal Green function for an inertial detector with the temperature\cite{BD} 
\be \label{T}
T_H=\d {a\hbar}{2\pi c k_B},
\ee
which leads to the conclusion that the uniformly acelerated detector in 
vacuum will detect a thermal bath of radiation at temperature $T_H$.  Thus,
the particle detectors moving along Eq.(\ref{traj}) will all
detect a thermal bath of radiation at the same temperature $T_H$.


\section{Concluding remarks}

A new kind of uniformly accelerated reference frame and the corresponding
coordinate system is constructed.  Unlike the M{\o}ller and
Rindler uniformly accelerated reference frames, the uniformly
accelerated reference frame has a single acceleration parameter
$a$. Although the distance between two observers standing in the
reference frame may change, the reference frame has the following
advantages:

 First of all, all observers with $X=const.$ in the reference
frame are on equal footing. In particular, each observer in the
reference frame has the same acceleration, and the reference frame
has spatial translation invariance. The frame is the relativistic
version of a uniformly accelerated reference in the sense of
Newtonian mechanics.   So, it is easy to be understood.  In
contrast, in the M{\o}ller and Rindler uniformly accelerated
reference frames there is no spatial translation invariance in the
direction of acceleration.  Instead, it possesses the boost
invariance.  The observers in the frame are not equivalent to each
other.  Therefore, they may be regarded as the uniformly accelerated
reference frame in a small nearby region
of a given observer.

Secondly, each observer in the new uniformly accelerated reference
frame has its own horizon (as shown in Fig.1).  It is something like the comoving
observers in Friedmann-Robertson-Walker universe when horizon
exists.  On the other hand, all standing observers have the same
horizon.  It is similar to de Sitter spacetime in the static
coordinate system.

Thirdly, each particle detector detects the same temperature in
the new uniformly accelerated reference frame.  It is again
something like the comoving observers in
Friedmann-Robertson-Walker universe, each of whom detects the same
temperature of cosmic microwave background.

\begin{figure}[hbt] \label{FIG:pieceacc}
\centerline{\includegraphics[scale=0.8]{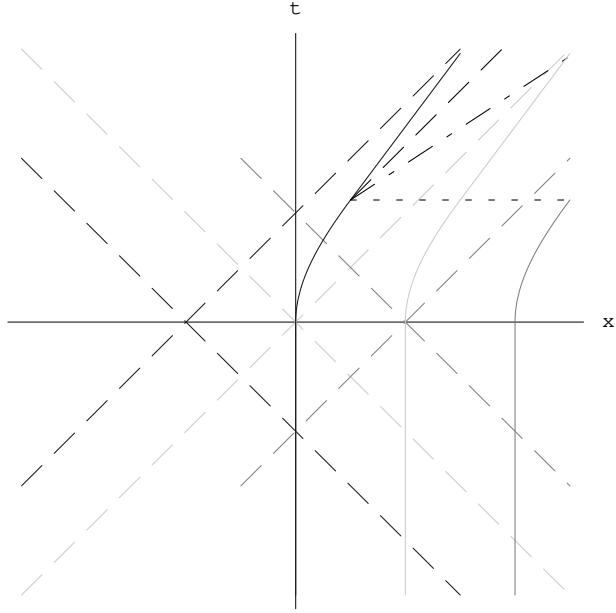}}
\caption{A reference frame which is first static, then uniformly accelerated, and finally moving at a constant velocity.  Real lines
are trajectories of observers.  Dashing lines in color are horizons,
and dashing line in black is the null cone when the acceleration
becomes zero.  Dash-dot line is the simultaneity surface of the
inertial reference frame which moves at the final velocity with
respect to the laboratory frame.  The horizontal dotted line is
the simultaneity surface of the laboratory (inertial) frame. }
\end{figure}

Fourthly, the reference frame may pick up a particular inertial
reference frame, the laboratory reference frame, in which the
distance of two standing observers keeps always unchanged and
observers (or particles) have the same instant velocity at any time
(The same temperature of different observers in a reference frame
may also be used to define a special, simultaneous hypersurface, because the thermal equilibrium is
closely related to the transitivity of simultaneity\cite{ZZ}). It
becomes clearer if a piecewise uniformly accelerated reference frame
is considered, which is first at rest in the laboratory, inertial
frame, and begins to be uniformly accelerated at $t=0$, and finally
settles down an inertial frame at $t=t_f$ with the final velocity
\be v_f=
\left . \d {dx} {dt} \right |_{t_f} = \frac {c^2t_f}{x(t_f)-X+\frac{c^2}{a}}=at_f
\left (1+ \frac{a^2t_f^2} {c^2}  \right )^{-1/2} \ee
with
respect to the laboratory frame, as shown in Fig.2. The reference
frame is more suitable than the M{\o}ller or Rindler uniformly
reference frame to apply the particles in a linear accelerator, in
which particles undergo the same acceleration, or the phonon in a
uniformly accelerated crystalloid.  The new frame can be used
to study the ``complicated fractional differences" due to acceleration,
which have been neglected in a small-enough accelerated frame\cite{MTW}.
The inertial
coordinates in the final inertial reference frame are related to
the coordinates for the laboratory reference frame
\be
\label{Inertframetransf}
t'&=&\d {t-t_f-v_fx/c^2}{(1-v_f^2/c^2)^{1/2}}, \nno \\
x'&=&\d {x-(c^2/a)[(1+a^2t_f^2/c^2)^{1/2}-1]-v_f t }{(1-v_f^2/c^2)^{1/2}}, \\
y'&=& y, \nno \\
z'&=&z. \nno
\ee
It is easy to see from Fig.2 that the distance between
the two observers with the same acceleration in the laboratory
frame keeps unchanged, while the distance in comoving frame
varies from time to time.

\section*{Acknowledgments}
This work is partly supported by NSFC under Grant Nos. 90403023,
90503002 and 10375087.

\end{document}